\documentclass[10pt]{article}

\makeatletter
\renewcommand{\fnum@figure}{\textbf{Fig.} \thefigure}
\makeatother

\usepackage{lineno,hyperref}
\usepackage{mwe}
\usepackage{caption}
\usepackage{subcaption}
\usepackage{color,soul}
\usepackage[round]{natbib} 
\usepackage{a4wide}
\usepackage{url}
\usepackage{float}
\usepackage{enumitem}
\usepackage{tabularx}
\usepackage{graphicx}
\usepackage{booktabs}
\usepackage{comment}

\title{YouTube and Science: Models for Research Impact}

\author{
    Abdul Rahman Shaikh$^1$\\
    ashaikh2@niu.edu
  \and
    Hamed Alhoori$^1$\\
    alhoori@niu.edu
    \and
    Maoyuan Sun$^1$\\
    smaoyuan@niu.edu
    }

\date{%
    $^1$Northern Illinois University\\%
}

\begin{document}

\maketitle

\begin{abstract}
Video communication has been rapidly increasing over the past decade, with YouTube providing a medium where users can post, discover, share, and react to videos. There has also been an increase in the number of videos citing research articles, especially since it has become relatively commonplace for academic conferences to require video submissions. However, the relationship between research articles and YouTube videos is not clear, and the purpose of the present paper is to address this issue. We created new datasets using YouTube videos and mentions of research articles on various online platforms. We found that most of the articles cited in the videos are related to medicine and biochemistry. We analyzed these datasets through statistical techniques and visualization, and built machine learning models to predict (1) whether a research article is cited in videos, (2) whether a research article cited in a video achieves a level of popularity, and (3) whether a video citing a research article becomes popular. The best models achieved F1 scores between 80\% and 94\%. According to our results, research articles mentioned in more tweets and news coverage have a higher chance of receiving video citations. We also found that video views are important for predicting citations and increasing research articles’ popularity and public engagement with science.
\end{abstract}


\section{Introduction}
\par
Social media platforms have seen tremendous growth and changed communication paradigms in the past decade, with information posted online within seconds and shared through multiple channels worldwide almost instantaneously. Immensely popular platforms such as Twitter, Facebook, and YouTube have shifted information-sharing on the internet from a largely one-way process of posting information to a proprietary website to one in which information becomes available rapidly through many websites and user accounts and in many forms and iterations, and in which responses to original content can be voiced and shared broadly based on a few manual clicks or via a preset automated process. This shift has seen trillions of posts, tweets, and video uploads on multiple platforms, which have had a far-reaching impact on most, if not all, areas of human endeavor—including scholarly research. Video communication has rapidly increased on social media sites such as YouTube, Instagram, Twitter, and Facebook. Launched in 2005, YouTube, in particular, has played a vital role in increasing video communication. In regard to scholarly work, research outcomes from this venue are increasingly being shared on several social media platforms. Further, it is by no means unusual for YouTube video descriptions to include citations of research articles, which could play an important role in disseminating research. In this paper, we analyze those video citations to better understand how research is being disseminated through such a new way of citation and the implications of this kind of sharing.
\par
In addition to the traditional citation-based analysis, it is necessary to track web-driven scholarly interactions and measure the impact of research beyond scholarly communities. Altmetrics or alternative metrics \citep{Cassidy} specifically track scholarly mentions on online platforms to analyze and measure the impact of scholarly products. Altmetrics track many different metrics, including, for example, users who have read or shared an article, even though those users may not formally cite it. Altmetrics could be used to measure the broader impact of a research article through many channels and even measure scholarly impact by predicting citations \citep{Thelwall_2018, AKELLA2021101128}. As a result, there is a movement toward digital libraries and publishers providing altmetrics on their websites. Several researchers have analyzed altmetrics and their potential for measuring societal impact. For example, \cite{BORNMANN2019325} used altmetrics data from the UK Research Excellence Framework (REF) in a case study designed to assess the validity of altmetrics and found that they have the potential to capture societal impact in relation to several distinct perspectives. In the present paper, we demonstrate our use of these social media metrics and features from altmetrics gathered from different platforms to further understand the role of YouTube video citations in the dissemination and societal impact of research.

Social media has become the primary source for real-time updates of various kinds of news worldwide, and YouTube is the principal site for video sharing \citep{Susarla2012}. \cite{Snelson2012} found that although YouTube is the most prolific online video-sharing platform, it is also the most under-researched social media platform. Every minute, more than 500 hours of video are uploaded to YouTube, which gets billions of views all over the world \citep{YouTubePress}. This popularity has resulted in YouTube ranking as the second most viewed website globally, outranked only by Google \citep{AlexaTopSites}. The platform’s usability and functionality allow a wide range of videos to be shared by users of all skill levels worldwide, who can then interact with others who like, dislike, and/or comment on the videos. Those posting content create a channel on YouTube as an organizing principle and showcase their video content, and others can choose to subscribe in order to receive updates when new videos are posted. Diverse fields of videos disseminating real-time information related to entertainment, health, music, sports, education, and news are uploaded by different channels. Recently, there has been an increase in the use of YouTube due to the dissemination of information related to the recent global pandemic, COVID-19 \citep{YouTubepop,YouTubepop2}.

The global COVID-19 pandemic has disrupted the lives of most people, requiring significant changes in many everyday practices. Social and professional activities worldwide have adapted new methods and processes to meet the challenges associated with these changes. In the academic world, as in other realms, annual conferences and society meetings are hosting events virtually \citep{falk2021international} in order to continue the exchanges that foster research progress in those venues. This unique experience has provided unexpected opportunities for the research community to reach wider audiences and improve diversity, equity, and novelty \citep{MichaelPrice,Bu,Thelwall-covid}. \cite{Beehler2020} hosted the 20th annual meeting of Interdisciplinary Nineteenth-Century Studies (INCS) online and shared the steps they took and the challenges they faced in hosting the conference in that context. Among many important points, the researchers asked the hosts to record the conference events and post the videos online to continue the discussion after the event. They also encouraged the presenters and other attendees to circulate appropriate hashtags on social media, share discussions, and broadcast the event’s news. Similarly, \cite{bonifati2020holding} shared their experience of hosting a joint conference online and provided a number of suggestions for doing so successfully, including implementing the practice of making videos of the presentations available to the public by posting them online after the conference. In such contexts, videos were mostly uploaded to the conference’s YouTube channel and often cited one or more articles in their description. In addition, many conferences require a video submission \citep{CHI2021, AAO2021}, and given that this is the case, there is a need to analyze and understand the characteristics and impact of these video citations within and beyond the scholarly community now that they are also available to the public.
\par
The idea of YouTube videos citing research literature is relatively new and has enormous potential to improve the visibility and dissemination of research, leading to greater societal and scholarly impact in the long term. However, there are far fewer studies on the relationship between research articles and scientific YouTube videos compared to the relationship between such articles and other social media platforms. To understand and analyze this impact, we consider the following research questions:

\smallskip

\begin{itemize}[label={}]
\item \textbf {RQ1.} What is the relationship between scientific YouTube videos and research articles? And what video categories and scientific subjects are most popular on YouTube?
  \item \textbf {RQ2.} Can we build machine learning models to predict the societal and scholarly impact related to YouTube videos? If so, what are the important features and characteristics? 
\end{itemize}

\smallskip

\par
To answer these questions, we analyzed the citations of research articles in videos on YouTube. We collected datasets from YouTube and Altmetric.com and used statistical and visualization techniques to understand the trends and discover patterns in the datasets. We combined the collected datasets and built machine learning models to predict three target variables. First, we built classification models to predict whether a research article has received a YouTube video citation. These video citations may influence the popularity of the research article, so it is vital to identify the features that are most important for contributing to the prediction of video mentions of research articles. Second, the prediction of citations of a research article is essential to assessing its scholarly impact. To understand the contribution of videos to this impact, we built our second machine learning model to predict citations of a research article using social media mentions and features from videos citing the article on YouTube. Third, the number of views of a video indicates the popularity of its content. To identify the social media features that are most important in attracting these views, we built our third machine learning model, which draws on social media features to predict the number of views for a video citing a research article on YouTube. In summary, our contributions include: 

\smallskip

\begin{enumerate}
   \item One of the first studies that examines the relationship between scientific YouTube videos and research articles. 
   \item An investigation of the societal and scholarly impact, characteristics, and popularity of YouTube videos through visualization, statistical techniques, and machine learning models. 
   \item  A novel dataset of research articles and YouTube videos that researchers can use to study the effects of scientific YouTube videos on the scientific community and the public.
\end{enumerate}

\section{Related Work}
\par
Users’ social interactions in the online world generate new forms of data that can be extracted and analyzed by developing new models to find undiscovered patterns beneficial to advances in research. The traditional analytical approach through scholarly citations limits measuring the impact within these scholarly boundaries. In contrast, altmetrics \citep{sud2014evaluating,BORNMANN2014895,Abdul-jcdl-2019} can measure the impact of research in a more diverse way across multiple platforms and can help us to understand the direct and indirect impact of scholarly research. On this point, according to \cite{Weller2015}, the proper use of altmetrics can help disseminate new scientific innovations to the public. Researchers have recently employed social media metrics to gauge public reaction to scientific findings~\citep{cole-2019-jcdl,cole-2020-group,Murtuza2022a,Murtuza2022b}.
\par
Now a feature of many people’s daily lives, YouTube has undeniably changed information-sharing worldwide. Since YouTube’s launch, researchers have studied the platform’s relationship with and impact on many aspects of daily life. For example, in a review of articles related to YouTube and health care, \cite{Madathil} found that health information is increasingly conveyed through YouTube, given that the platform offers a setting for users worldwide to upload, view, and communicate health information. Further, as an educational tool in nursing education \citep{Agazio2009,JOHNSTON2018151}, the platform has been used to stimulate active learning on the part of students and enhance health care learning. \cite{Chtouki} found that YouTube videos can be a useful source of free educational content that improves student performance. In a study designed to assess the effectiveness of YouTube videos on a given anatomy problem, \cite{Jaffar2012} found that 98\% of the 91 medical students used YouTube extensively, and 92\% agreed it helped them learn about anatomy. \cite{June2014} conducted research with a sample of 50 students to assess their critical-thinking skills and interactive activities while using videos on YouTube. According to the results, YouTube has vast potential as a learning tool because it enhances the students’ experience and improves their critical-thinking skills.
\par
Due to the massive quantity of content on YouTube, it is generally difficult for most contributors to reach a vast audience. Therefore, there is no guarantee that posting content will create an impact. For this reason, the relative popularity of videos has become a research focus with the goal of determining the features that are principally responsible for a video achieving a high level of popularity. According to \cite{Susarla2012}, older YouTube accounts with more videos posted have a greater chance of having an impact than do newer accounts with fewer videos. In an assessment of the effect of content-agnostic factors on video popularity, \cite{Borghol_2012} found that views of videos shared previously by the uploader and video age are critical factors in determining video popularity. \cite{Figueiredo2014} used Amazon Mechanical Turk to evaluate YouTube videos and found that high-quality content resulted in tremendous popularity. To analyze the factors affecting the popularity of videos focused on science communication on YouTube, \cite{Welbourne2016} studied 390 videos of this kind uploaded to 39 channels on the platform and extracted popularity metrics such as view count, comment count, subscriber count, share count, and rating. They found that user-generated content was more popular than professionally generated content. \cite{Broderson} examined more than 20 million YouTube videos to determine the relationship between video popularity and geographic locality by analyzing the views from a spatial locality instead of a global locality. They found that the videos have a strong geographic interest, and around 50\% of the videos had gained more than 70\% of their views from a single region. \cite{Khan2014} analyzed top viral videos by building an empirical model to understand how videos achieve virality and the relationship between different aspects such as social and non-social capital. They found that along with view count, offline social capital and network dynamics are the strongest contributors to virality.
\par
Several models have been proposed to predict the popularity of videos on YouTube. \cite{Pinto2013} proposed two regression models to predict the popularity of videos using two YouTube video datasets. They found that variables such as the number of comments, ratings, and users who favorited the video are usually positively correlated with the number of views and do not help the regression model. Instead, information related to the user who posted the video and the subscriber count proved most useful. 
\cite{Hovden} conducted an early study that investigated the productivity and impact of the top video channels on YouTube by applying h-index and g-index bibliometrics. Music video-based channels had a high impact appearing in the g-index rankings, whereas video blogs and mini-shows ranked top in the h-index rankings. They found that these metrics were best if used to compare channels of a related field.
\cite{Yu2015} collected 172,000 videos from YouTube and proposed popularity phases to describe a YouTube video’s lifecycle. They found multiple stages of popularity with increases and decreases over several months for most videos, with phases related to content and popularity. \cite{Ma2017} proposed a Lifetime Aware Regression Model (LARM) to predict long-term video popularity using early accessible features such as views, likes, dislikes, comments, and video categories. They used two YouTube datasets to validate their model and found that it outperformed other baselines by up to 20\% in terms of the prediction error reduction rate. \cite{Trzcinski2017} proposed a Support Vector Regression model with Gaussian radial basis functions to predict the popularity of videos on YouTube and Facebook. They found that social features are more strongly associated with video popularity predictions than visual features.
\par
Most of the previous research on YouTube focuses on the potential use of the platform as a tool for communicating information on health, politics, or science and as a means for providing education for students to facilitate learning. Researchers have built models to predict the popularity of videos on YouTube through metadata about the videos, and they have conducted surveys and performed literature reviews to understand users’ behavior. Most research to date has not explored the connection between YouTube videos and research outcomes. Based on our research, the present paper is one of the first in which the relationship between research articles and YouTube videos is analyzed through a range of social media features.

\section{Methodology}

\subsection{Data Collection and Preprocessing} 
The data for this study came from two sources: Altmetric.com and YouTube. We analyzed these datasets individually and then combined them to form final datasets consisting of several features, which consist of social media mentions of the research articles and the metadata of the YouTube videos citing the research articles, as shown in Table \ref{tab:features1} and Table \ref{tab:features2}. Using these datasets, we applied visualization techniques and statistical models for analysis and built machine learning models to predict important features in the dataset.

\begin{figure}[htbp]
  \centering
  \includegraphics[width=\linewidth]{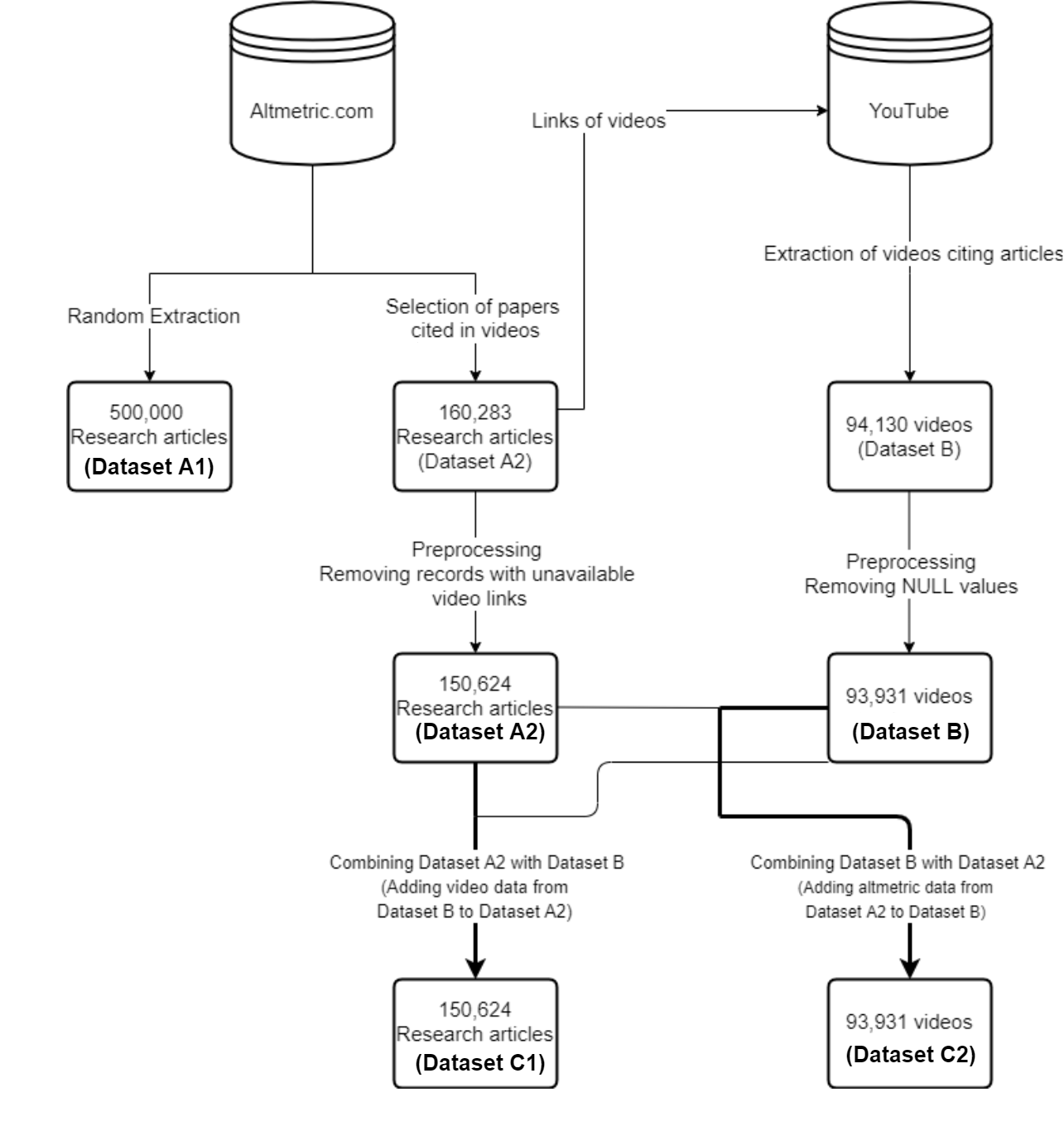}
  \caption{Data collection process}
  \label{fig:dataprocess}  
\end{figure}
\par
The data collection process for this research is shown in Figure \ref{fig:dataprocess}. We extracted 500,000 random research articles from the Altmetric API, which includes metadata of the research articles and mentions of research outputs on various social media networks. We named this dataset A1; it includes research articles cited in videos as well as research articles not cited in videos. Descriptions of the features in this dataset are presented in Table \ref{tab:features1}.
\begin{table}
\centering
\caption{Altmetric features of datasets A1 and A2}
\label{tab:features1}
\newlength\q
\setlength\q{\dimexpr .75\textwidth -0.2\tabcolsep}
\noindent\begin{tabular}{|l|p{\q}|}
\toprule
\textbf{Feature} & \textbf{Feature Description}\\
\midrule
Altmetric ID&Unique ID for each research article\\\hline
Title&Title of a research article\\\hline
Publication Date&Publication date of a research article\\\hline
Mendeley Readers&Number of times a reference to a research article is archived on Mendeley\\\hline
Scopus Subjects&Subjects of the research article\\\hline
News&Number of times a research article is mentioned in news contexts\\\hline
Twitter&Number of times an article has been tweeted on Twitter\\\hline
Facebook&Number of times an article is shared on Facebook\\\hline
Policy&Number of times an article is shared in policy documents\\\hline
GooglePlus&Number of times an article is shared on Google Plus\\\hline
Reddit&Number of times an article is mentioned on Reddit\\\hline
Blogs&Number of times an article is mentioned or featured in blogs\\\hline
Patent&Number of times an article is mentioned or featured in patents\\\hline
Wikipedia&Number of times an article is cited on Wikipedia\\\hline
Video Citations&Number of times an article is mentioned in the description of videos (i.e., cited) on YouTube\\\hline
Citation&Number of times a research article is cited based on Dimensions.ai\\\hline
YouTube Links&List of YouTube video links citing the publication (dataset A2)\\\hline
YouTube Citation&Binary variable indicating whether or not the publication is cited on YouTube. This is the target variable for Model 1 and is used in dataset A1\\\hline
Scholarly Citation&Binary variable indicating whether the publication has more citations than the median number of citations, which is 27 for all publications in dataset A2. This is the target variable for Model 2 and is used in dataset A2\\
\bottomrule
\end{tabular}
\end{table}
\par
The Altmetric dataset provided the number of videos in which an article was cited with links to those videos. We checked all those links and verified they belonged to the YouTube platform. Each video on YouTube cited research articles through the description of the video uploaded to the channel. If the value of the feature Video was greater than 0, then we set the target variable to 1. Otherwise, we set the target variable to 0. We named this binary feature, which served as a target variable, “YouTube Citation.”
\par
To further explore articles cited in videos, we selected all the research articles in altmetric.com that are cited by at least one video on YouTube. 
Using the Altmetric API, we collected research articles cited by video along with the list of video links citing the publication. 
We found a total of 160,283 research articles cited in videos on YouTube. 
We extracted these articles and created dataset A2, which comprises Altmetric IDs (a unique ID for each research article), metadata of those research articles, and post counts from various social media platforms, as shown in Table \ref{tab:features1}. The features are the same for both datasets with the exception of the target variable, which is “YouTube Citation” for A1 and “Scholarly Citation” for A2. Of the original 160,283 research articles in dataset A2, 9,659 records did not have valid video links. We, therefore, removed these records, resulting in a final count of 150,624. The Scopus subjects of the research articles were split to include only the major subjects, and the sub-subjects were removed. A total of 22 unique major subjects were analyzed further, as described in Section 3.4. For 21,905 research articles, no major subject or sub-subject was defined. We, therefore, converted these empty values to “Other” subjects.
\par
We then created dataset B through the video links collected in dataset A2 under “YouTube Links” by extracting the videos’ metadata through the YouTube API. There were 94,130 unique video links in dataset B, of which 199 video links were unavailable, leaving 93,931 videos with available links and containing the YouTube video ID or link, the title of the video, views, likes, dislikes, description, the number of comment counts, and cited IDs of Altmetric articles. It is important to note that an article could be cited in more than one video and that one video could cite more than one article. Dataset B contains only the metadata of videos citing research articles, as shown in Table \ref{tab:features2}.

\begin{table}
\caption{Features from YouTube}
\label{tab:features2}
\newlength\qq
\setlength\qq{\dimexpr .75\textwidth -0.2\tabcolsep}
\noindent\begin{tabular}{|l|p{\qq}|}
\toprule
\textbf{Feature} & \textbf{Feature Description}\\
\midrule
Link&YouTube unique link or ID of video\\\hline
Cited ID&Altmetric IDs of cited research articles\\\hline
Title&Title of YouTube video\\\hline
Views&Total views of a YouTube video\\\hline
Likes&Total likes of a YouTube video\\\hline
Dislikes&Total dislikes of a YouTube video\\\hline
SubNo&Total subscriber count of the channel posting the video\\\hline
Pubdate&Publication date of the video\\\hline
Description&Description of the video as provided by the channel\\\hline
Video Category&Category of the video as stated by the uploader\\\hline
Comments&Number of comments posted on the YouTube video\\\hline
Video Views&Binary variable indicating whether or not the video has more than the median number of views, which is 1,139 for all the videos in dataset B. This is the target variable for Model 3.\\
\bottomrule
\end{tabular}
\end{table}

\par
In dataset B, of the 93,931 videos, 11,710 had no “Likes” values, and 2,553 had null values in the Likes feature, which both were converted to 0 “Likes .” Similarly, 42,673 videos had no “Dislikes,” and 2,554 null values in the Dislike feature were converted to 0 “Dislikes.” In addition, 64 videos had no “Views” in the Views feature and were converted to 0 “Views.” The channel subscriber count is presented in a textual form such as “1.2M” or “330K” and was converted appropriately to integer values. Features such as “Views,” “Likes,” “Dislikes,” “Subno,” and “Comments” were converted into integers from strings. For the feature “Pubdate,” there were 4 null values and 2,823 values that included “Premiered on Date,” which were changed to a date format from a string. We checked the “Category” feature of videos and found 15 categories out of the 29 major categories on YouTube \cite{TechPost}. There were 2,126 “Category” values different from the 15 categories, which we changed to the “Undefined” category.

\subsection{Approach}

In our approach to answering RQ1, we use dataset A1 to build our first machine learning model to predict video citation. Analyzing the important features of the first model helped us further explore the relationship between videos and articles and is also an indirect indicator of popularity. The analysis of dataset A1 revealed a class-imbalance problem while building the machine learning model, as 84\% of research articles were not cited in videos. This shows that the majority of articles are not cited in videos, but popular articles might be cited in videos. For the binary feature “YouTube Citation," the class 0 value denoting a research article not cited in videos contained 419,886 records, whereas class 1 indicating a research article cited in videos had 80,114 records.
\par
In a real-world classification of any given dataset, the classes have a high chance of imbalance. Thus, the classes are not relatively equal in number, i.e., one of the class values is higher than the others. However, there are several techniques to resolve this issue. A class-imbalance problem can be resolved by either oversampling the minority class or undersampling the majority class. The goal was to process the imbalanced data before feeding it into the classifier and, in this way, overcome the fact that the classifier is more sensitive to the majority class and less sensitive to the minority class, resulting in the prediction of the majority class. A limitation of oversampling a minority class in an imbalanced dataset is that it could lead to overfitting, as the existing values are copied in this technique. The concern with undersampling is that essential data can be missed. 
\par
We applied a hybrid method to dataset A1, combining random oversampling and undersampling techniques. In the case of oversampling applied to the dataset, the minority class samples were increased to equal 0.5 of the majority class samples. After oversampling the dataset, we applied undersampling in which the majority class samples were downsampled to equal 0.8 of the minority class samples. The class label counts from the dataset that resulted from applying the hybrid method were as follows: 262,428 for class 0, which denotes a research article not cited in videos, and 209,943 for class 1, which denotes a research article. This reduced the effect of the class-imbalance problem for dataset A1. We then applied all the classification models to the resulting dataset A1 to determine which model performed best.
\par
To answer RQ2, we combined dataset B with dataset A2 to create datasets C1 and C2, our final datasets, which we used in building models to predict article citations and video views, respectively. The second model used dataset C1 and the features utilized are shown in Table \ref{tab:overview}. The feature “Scholarly Citation" described in Table \ref{tab:features1} is the target variable for the second model, which represents article citations. Citation analysis of articles is one of the indicators of scholarly impact and can help identify the important features related to scholarly impact through videos on YouTube. The final model is built using dataset C2 and the features utilized are shown in Table \ref{tab:overview}. The feature “Video views" is the target feature for the third model described in Table \ref{tab:features2}. Views of a YouTube video are associated with popularity and are a potential indicator of societal impact. Important features of the model can help in examining features related to societal impact through views of videos. A description of all the datasets and available features is shown in Table \ref{tab:Description}.

\begin{table}
\footnotesize
\caption{Metadata about the Datasets utilized in our work. ( * =features averaged over all research articles; ** = features averaged over all videos citing research articles).}
\label{tab:Description}
\newlength\qqq
\setlength\qqq{\dimexpr .55\textwidth -0.2\tabcolsep}
\setlength\qq{\dimexpr .20\textwidth -0.2\tabcolsep}
\noindent\begin{tabularx}{\textwidth}{|c|p{\dimexpr.25\linewidth-2\tabcolsep-1.3333\arrayrulewidth}|X|}
\toprule
\textbf{Dataset}&\textbf{Description}&\textbf{Features}\\
\midrule
A1& Random 500,000 Research articles from Altmetric.com & Altmetric ID, Title, Publication Date, Scopus Subjects, Video Citation, Citation, News, Twitter, Facebook, Policy, Google+, Reddit, Mendeley Readers, Patent, Blogs, Wikipedia, YouTube Citation \\\hline
A2& All 150,624 Research articles from Altmetric.com cited in available videos (YouTube) & Altmetric ID, Title, Publication Date, Scopus Subjects, Video Citation, Citation, News, Twitter, Facebook, Policy, Google+, Reddit, Mendeley Readers, Patent, Blogs, Wikipedia, YouTube Links, Scholarly Citation \\\hline
B& All 93,931 available videos citing research articles & Link, Title, Views, Likes, Dislikes, SubNo, Pubdate, Description, Video Category, Comments, Video Views, Cited Altmetric IDs \\\hline
C1& 150,624 Research articles with metadata of all videos citing articles & Altmetric ID, Title, Publication Date, Scopus Subjects, Video Citation, Citation, News, Twitter, Facebook, Policy, Google+, Reddit, Mendeley Readers, Scholarly Citation, Patent, Blogs, Wikipedia, Video Views*, Video Likes*, Video Dislikes*, Video SubNo*, Video Comments* \\\hline
C2& 93,931 YouTube videos with metadata of articles cited by each video & Link, Title, Views, Likes, Dislikes, SubNo, Pubdate, Description, Video Category, Comments, Video Views, Video Citation**, Citation**, News**, Twitter**, Facebook**, Policy**, Google+**, Reddit**, Mendeley Readers**, Patent**, Blogs**, Wikipedia** \\
\bottomrule
\end{tabularx}
\end{table}

To determine the importance of video citations for popularity or increasing the citations of a research article, we combined datasets A2 and B to form dataset C1, which consists of all the research articles cited by a video on YouTube with the links to the videos and the metadata of those videos citing that research article. Dataset C1 was created by adding video metadata from dataset B to its link present in the Video links of dataset A2. The final dataset C1 comprises 150,624 research articles cited by 93,931 unique video links. The features of the videos in dataset C1 have multiple values, as each research article had video citations ranging from 1 to 814 videos. Numerous videos can cite an article, and each video has different values for its features. Therefore, in building our models, we averaged the values of all the features of the YouTube videos citing a research article. For example, for a research article cited by three videos with respective view counts of 35634, 2733, and 1, we determined a single average value of 12,794.34.
\par
To further analyze the YouTube videos citing research articles, we combined datasets B and A2 to create dataset C2, which consists of 93,931 YouTube videos citing 150,624 research articles, along with the metadata of the videos and the altmetric information of the research articles cited in the video descriptions. Dataset C2 was created by adding altmetric data from dataset A2 to its altmetric ID present in the feature for the cited ID in dataset B. The altmetrics features in dataset C2 have multiple values, as each YouTube video cited numerous research articles ranging from 1 to 82. Therefore, we averaged the numerical features for the research articles. For example, if a video cites five research articles and those articles had Twitter mentions such as 80, 63, 31, 27, 15, they were averaged to a single value of 43.2. It is important to note that in dataset C1, an article could be cited by multiple videos, whereas in dataset C2, a single video could cite numerous articles. For data exploration, we used datasets A1, A2, B, and C1, whereas in building the machine learning models, we used only datasets A1, C1, and C2. The resulting datasets are publicly available as a comma-separated-value file (CSV)\footnote[1]{https://zenodo.org/record/4691941}.

The datasets used in building the models differ from one another and are presented in Table \ref{tab:overview}, along with features utilized in each model and the target variable. Given its correlation with Mendeley Readers, we removed Citations from dataset A1. For Model 3, in dataset C2, Citations and Blogs were removed as they are correlated with other features. For each model, we used Scikit-learn \citep{pedregosa11a} to implement classification models such as Bernoulli Naive Bayes, Random Forest, Decision Tree, and K-Nearest Neighbors (KNN). For the machine learning models, we applied k-fold cross-validation to the dataset and average evaluation metrics over 20 folds. To evaluate each model, we used the metrics precision, recall, F1-score, and accuracy. The four models are described along with the evaluation metrics. We report the essential features in building all the models to identify the importance of those attributes in relation to the target variables and our research questions. 

\begin{table}
\caption{Datasets, features, and target variable considered for all models ( * =features averaged over all research articles; ** = features averaged over all videos citing research articles).}
\label{tab:overview}
\setlength\qqq{\dimexpr .55\textwidth -0.2\tabcolsep}
\noindent\begin{tabular}{|c|c|p{\qqq}|c|}
\toprule
\textbf{Model}&\textbf{Dataset}&\textbf{Features}&\textbf{Target}\\
\midrule
1.&A1&News, Twitter, Facebook, Policy, Google+, Reddit, Mendeley Readers, Patent, Blogs, Wikipedia &YouTube Citation \\\hline
2.&C1&Mendeley Readers, Twitter, Facebook, Policy, Google+, Patent, Wikipedia, Video Citations, Video views**, Subno**, Comments** &Scholarly Citation \\\hline
3.&C2&Mendeley Readers*, Video Mentions*, Facebook*, Google+*, Twitter*, Wikipedia*, Policy*, Patent*, Reddit*, Video Category, SubNo &Video Views\\
\bottomrule
\end{tabular}
\end{table}

\section{Results}
We report the results of our work in two parts. The first part describes the exploration of the dataset, which helps answer RQ1. The second part deals with building machine learning models to predict different metrics for answering RQ2. We provide an analysis of the datasets through visualization and statistical techniques in the first part and then build classification models for the second part.

\subsection{Data Exploration}

\begin{figure}[htbp]
  \centering
  \includegraphics[width=\linewidth]{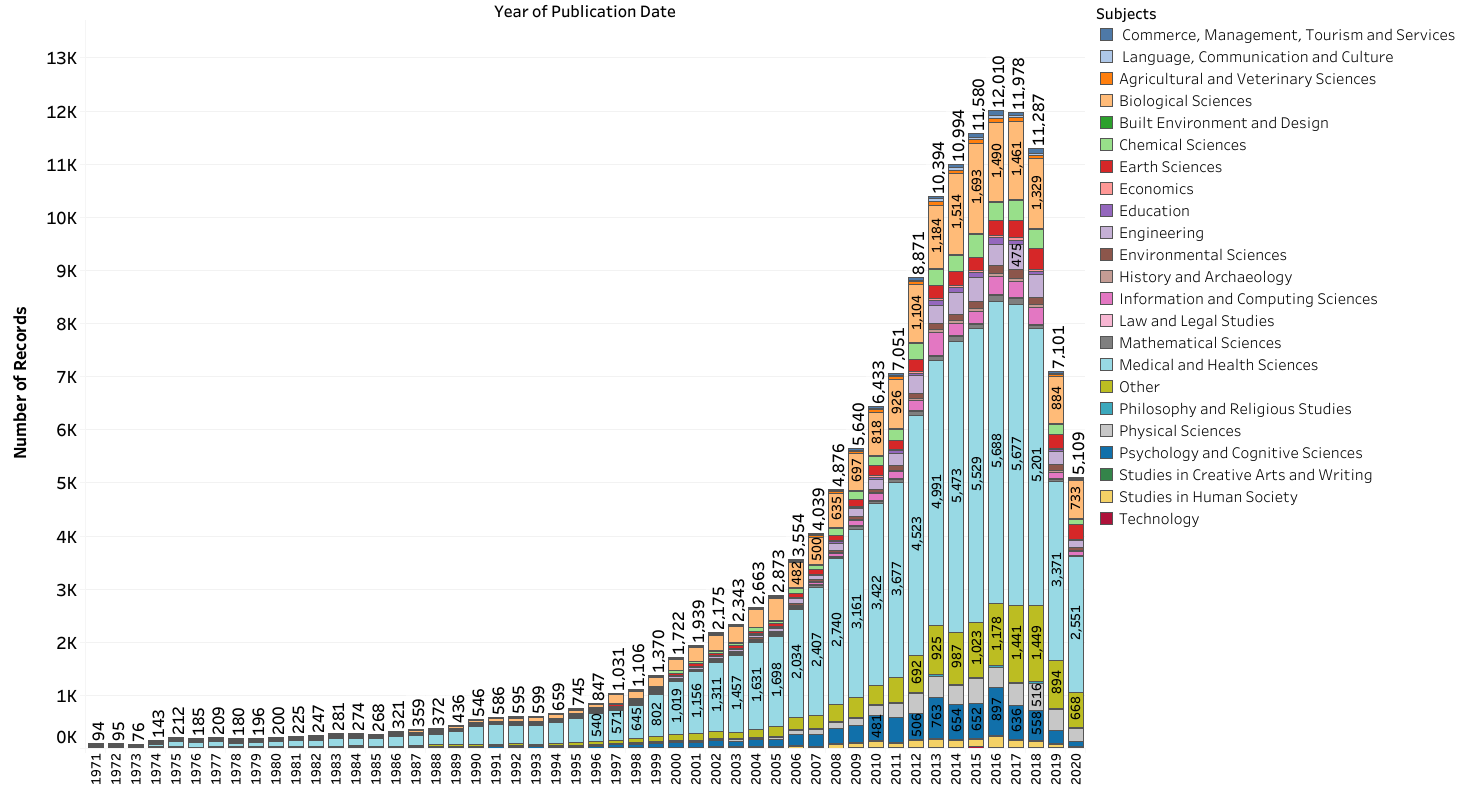}
  \caption{Number of records of Scopus subjects by year in dataset A2}    \label{fig:datanalysis1}
\end{figure}
\par
\begin{figure}[htbp]
  \centering
  \includegraphics[width=\linewidth]{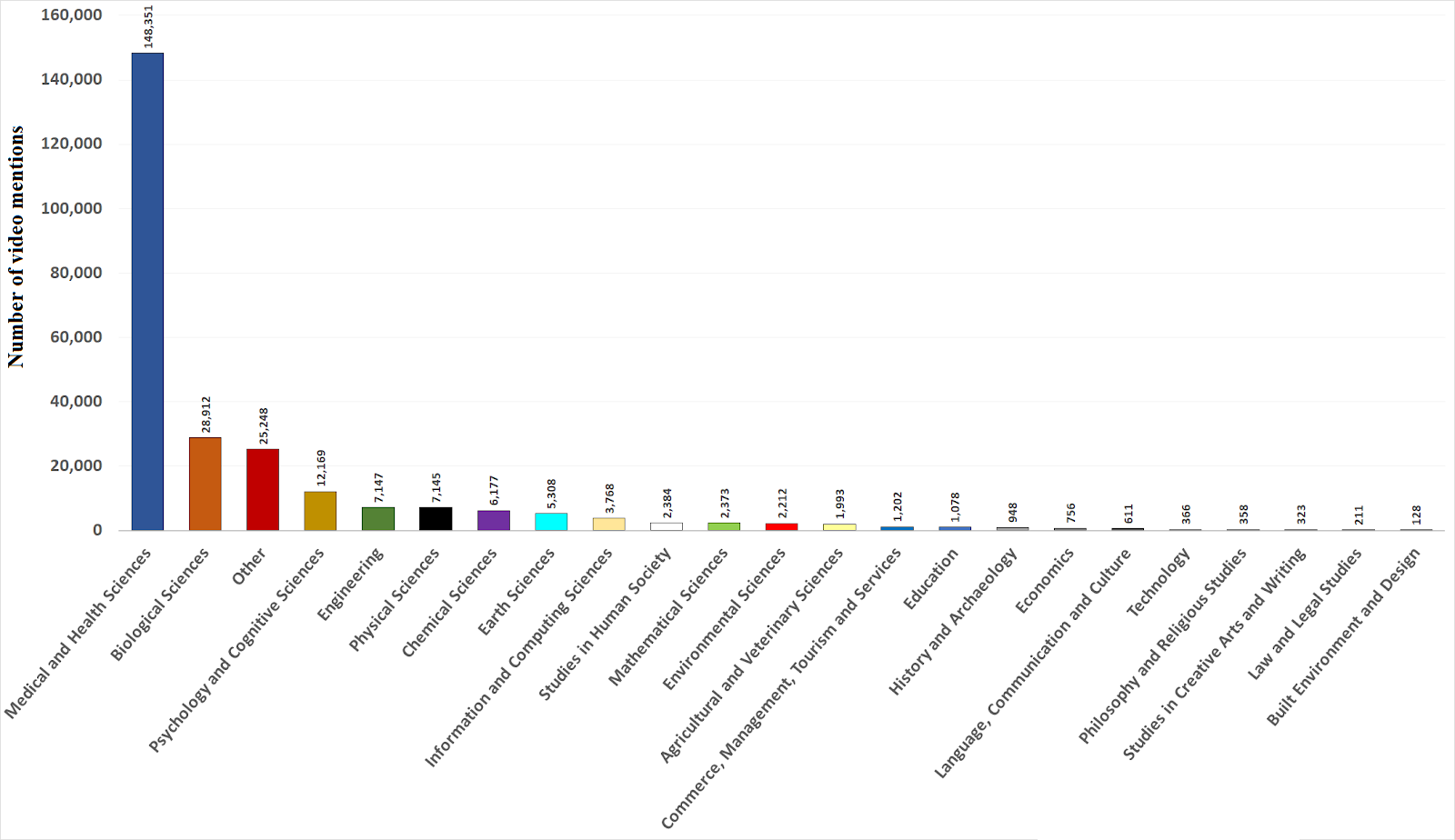}
  \caption{Number of video mentions for Scopus subjects in dataset A2}
  \label{fig:datanalysis2}
\end{figure}

To explore the second part of RQ1 and identify the popular subjects of articles cited in videos, we analyzed the research articles in dataset A2 and found that an increasing number of research articles have been cited in videos in the last two decades (Figure \ref{fig:datanalysis1}). The articles most cited in videos each year were the “Medical and Health Sciences” Scopus subject. As the data was collected at the end of 2020, we can see a drop in the number of research articles cited in videos in 2019 and 2020 compared to previous years. This could mean that articles tend not to be cited in videos shortly after publication, with citations of this kind taking time to accrue. We also found that research articles with the Scopus subjects Medical and Health Sciences, Biological Sciences, and Psychology and Cognitive Sciences had the highest number of video mentions in comparison to other subjects (Figure \ref{fig:datanalysis2}). This indicates that most of the research articles cited in the videos were related to medicine and biochemistry. The citation count for these subjects was also higher than for other subjects.

\begin{figure}[htbp]
  \centering
  \includegraphics[width=\linewidth]{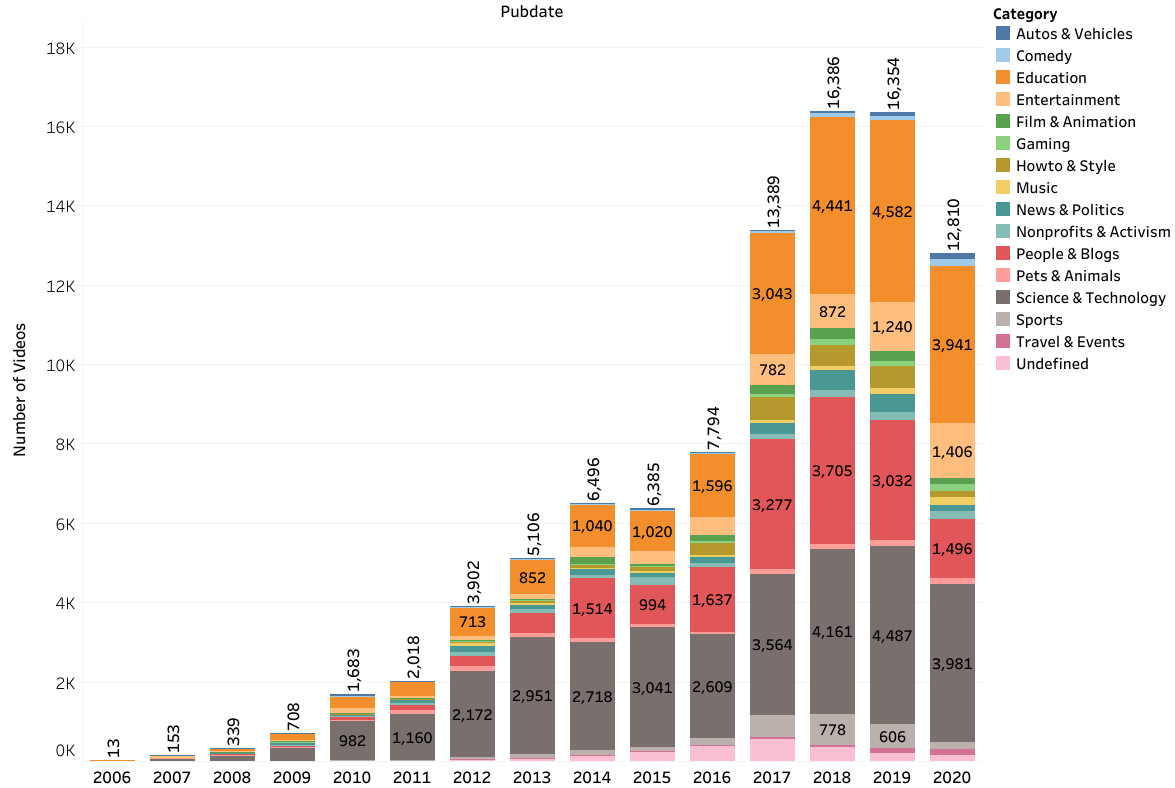}
  \caption{Number of YouTube videos in each category per year}
  \label{fig:datanalysis3}
\end{figure}
\par

\begin{figure}[htbp]
  \centering
  \includegraphics[width=\linewidth]{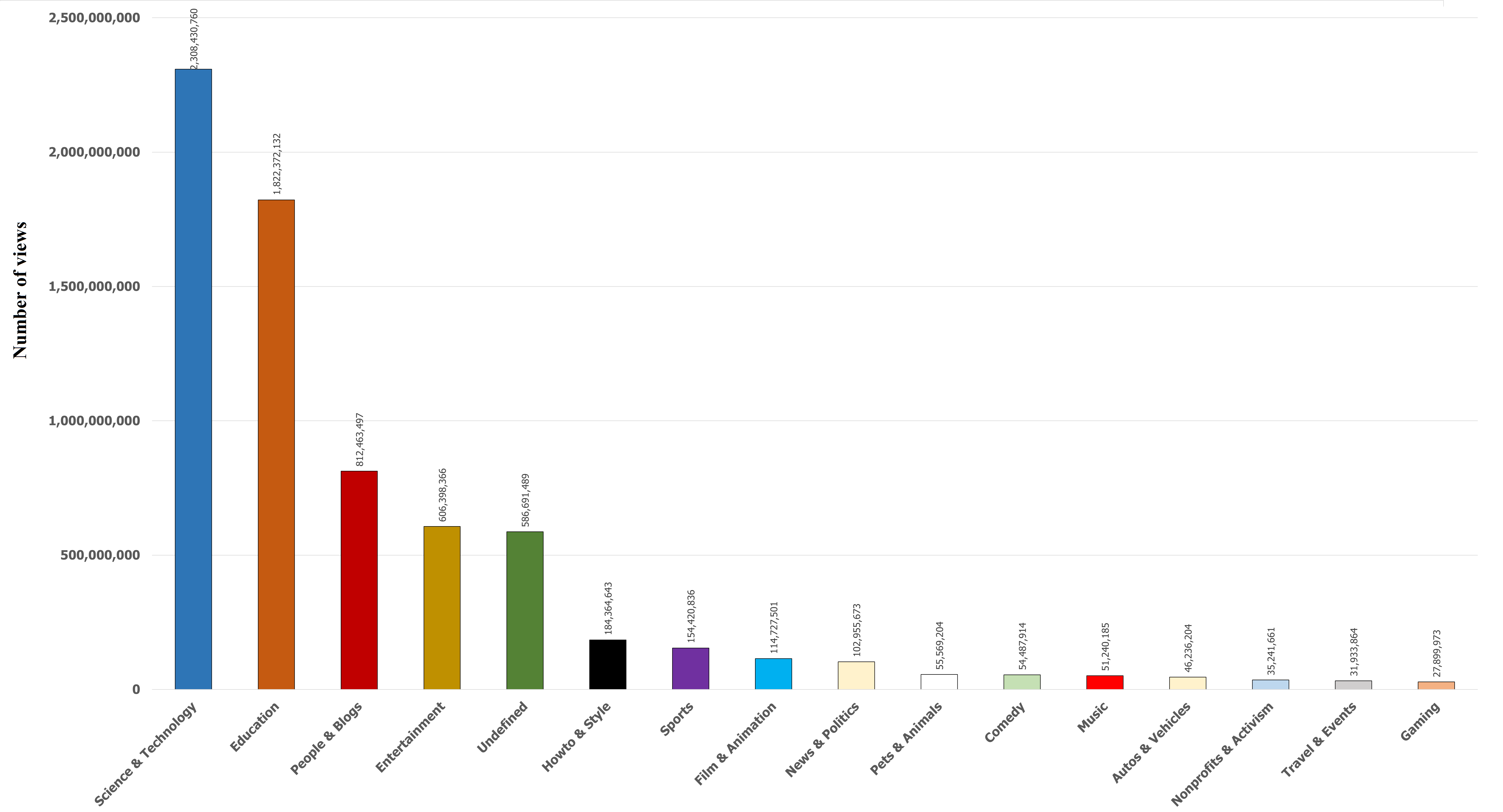}
  \caption{Video views of YouTube categories}
  \label{fig:datanalysis4}
\end{figure}

The popular video categories on YouTube citing research articles can be seen in Figure \ref{fig:datanalysis3}. We can see that an increasing number of videos have cited research articles in the last decade. The majority of the videos citing research articles are categorized as Education, Science \& Technology, or People \& Blogs. It is also worth noting that People \& Blogs was the third-highest category of videos to cite research articles, even though most of the research articles cited belong to Medicine \& Biochemistry. Figure \ref{fig:datanalysis4} shows the number of views of YouTube videos in dataset B based on the video category. The most-viewed videos citing research articles are categorized as Science \& Technology and Education due to the high number of these videos in the dataset. Even though fewer videos were categorized as Entertainment in dataset B, this category garnered the fourth-highest view count.

\begin{figure}[htbp]
  \centering
  \includegraphics[width=\linewidth]{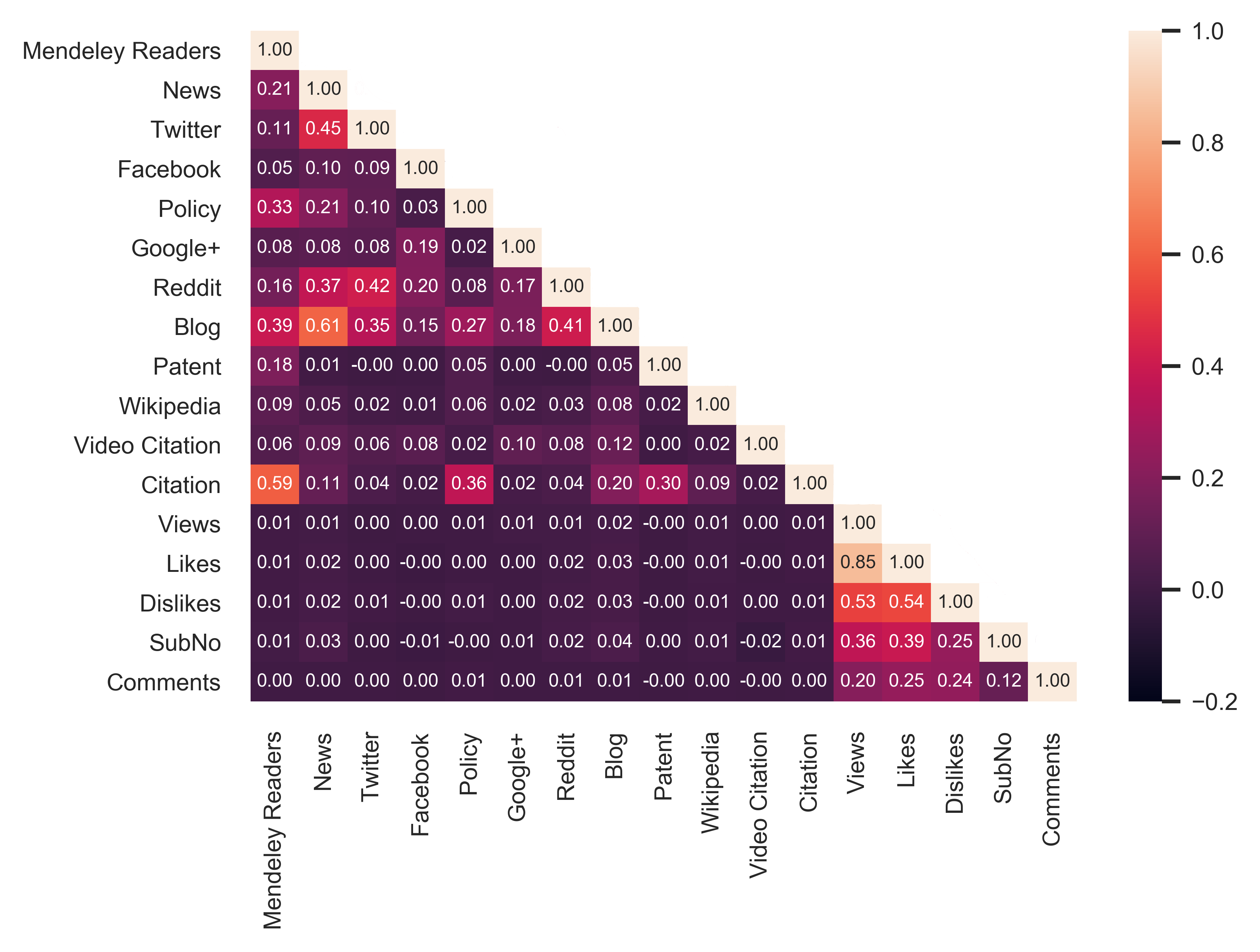}
  \caption{Correlation matrix for all the numerical features in dataset C1}
  \label{fig:datanalysis5}
\end{figure}

To explore the relationship between articles and video features, we plotted a correlation matrix between the features to answer RQ1. Based on the correlation matrix of the features in dataset C1, we can observe that Likes and Dislikes are highly correlated with Views (Figure \ref{fig:datanalysis5}). 
To reduce complexity while building our machine learning models, we removed highly correlated features, i.e., features with a correlation greater than 0.5. Thus, we removed Likes and Dislikes as they had a correlation of 0.85 and 0.53, respectively, keeping only video Views as the sole feature relating to the videos. We also removed Blog as it had a 0.61 correlation with News, and Citation was also removed due to its correlation of 0.59 with Mendeley Readers.

\subsection{Building Models}

\begin{figure}[htbp]
  \centering
  \includegraphics[width=\linewidth]{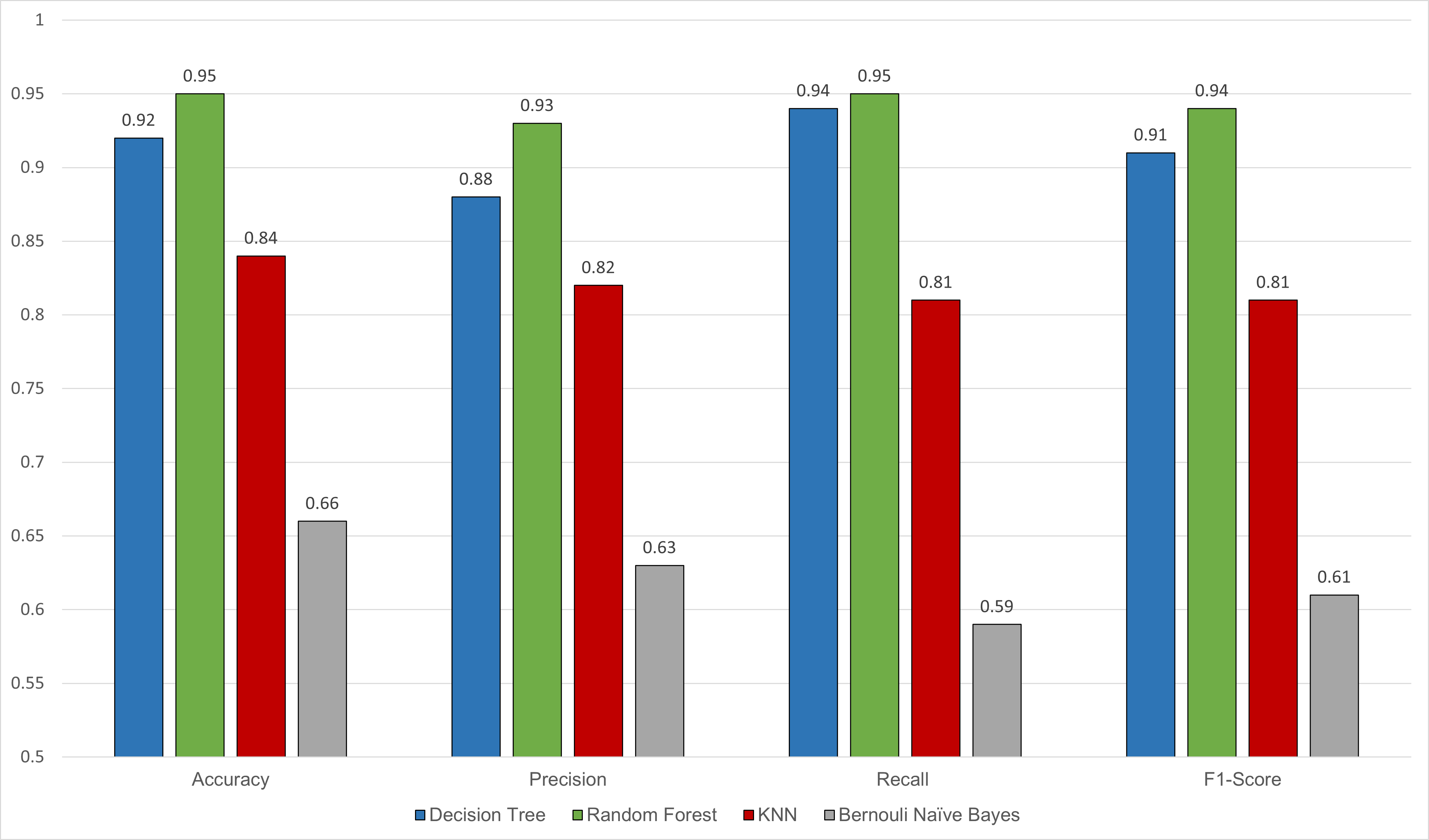}
  \caption{Classification results for Model 1}
  \label{fig:Results1}
\end{figure}

\begin{figure}[htbp]
  \centering
  \includegraphics[width=\linewidth]{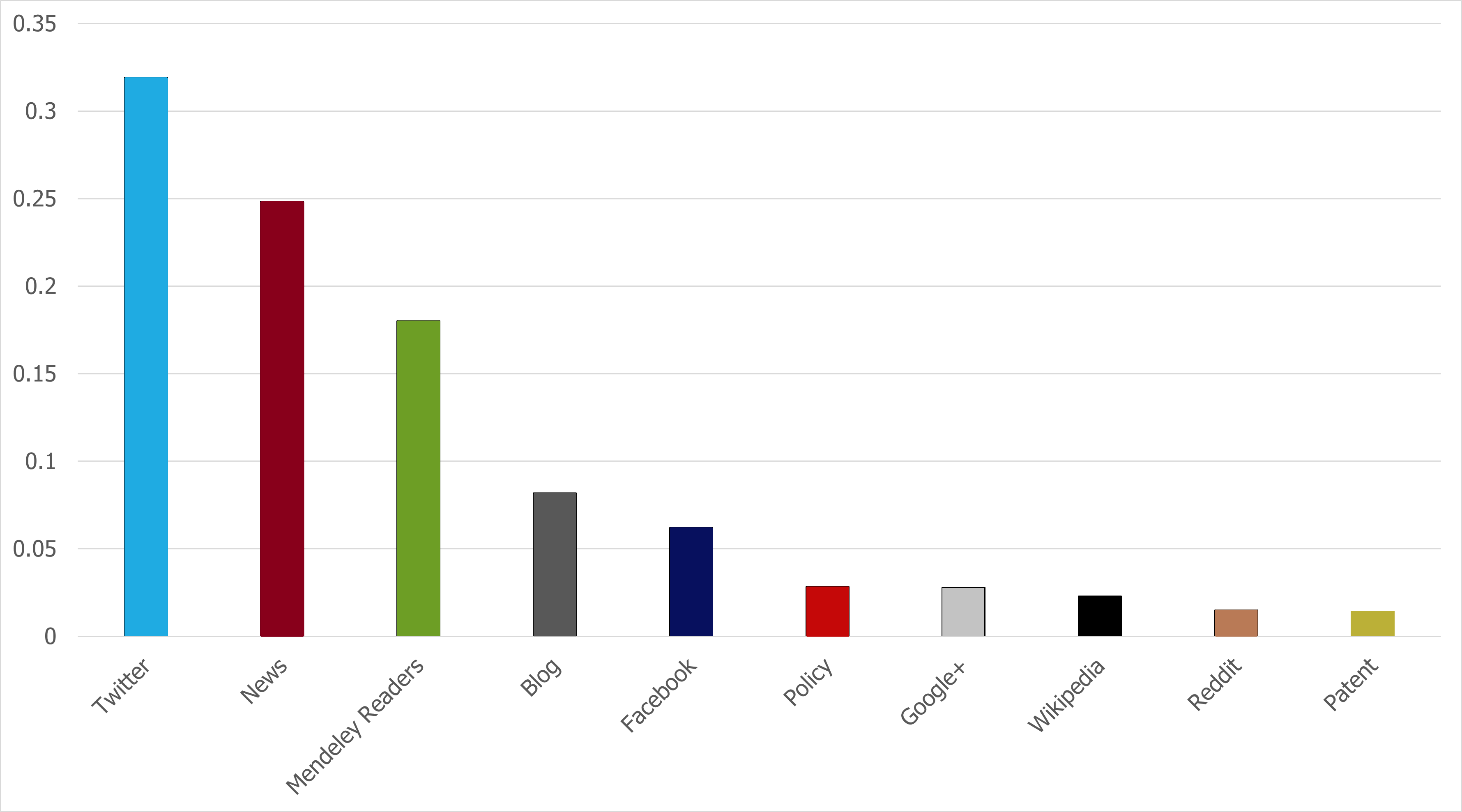}
  \caption{Important features for the Random Forest classifier for Model 1}
  \label{fig:Feature1}
\end{figure}
To explore the relationship between articles and videos as well as predict the impact of popularity, we predicted whether an article was cited in at least one video on YouTube. This identified important social media features of an article influencing YouTube citation. In the first model, Random Forest yielded the best results, as shown in Figure \ref{fig:Results1}, achieving an accuracy level of 0.95 and an F1-score of 0.94. Decision Tree performed second best with an accuracy of 0.92 and an F1-score of 0.91. Figure \ref{fig:Feature1} shows the feature importance of the Random Forest classifier in predicting video citations. Twitter and News were the most important features, followed by Mendeley Readers and Blog.

\begin{figure}[htbp]
  \centering
  \includegraphics[width=\linewidth]{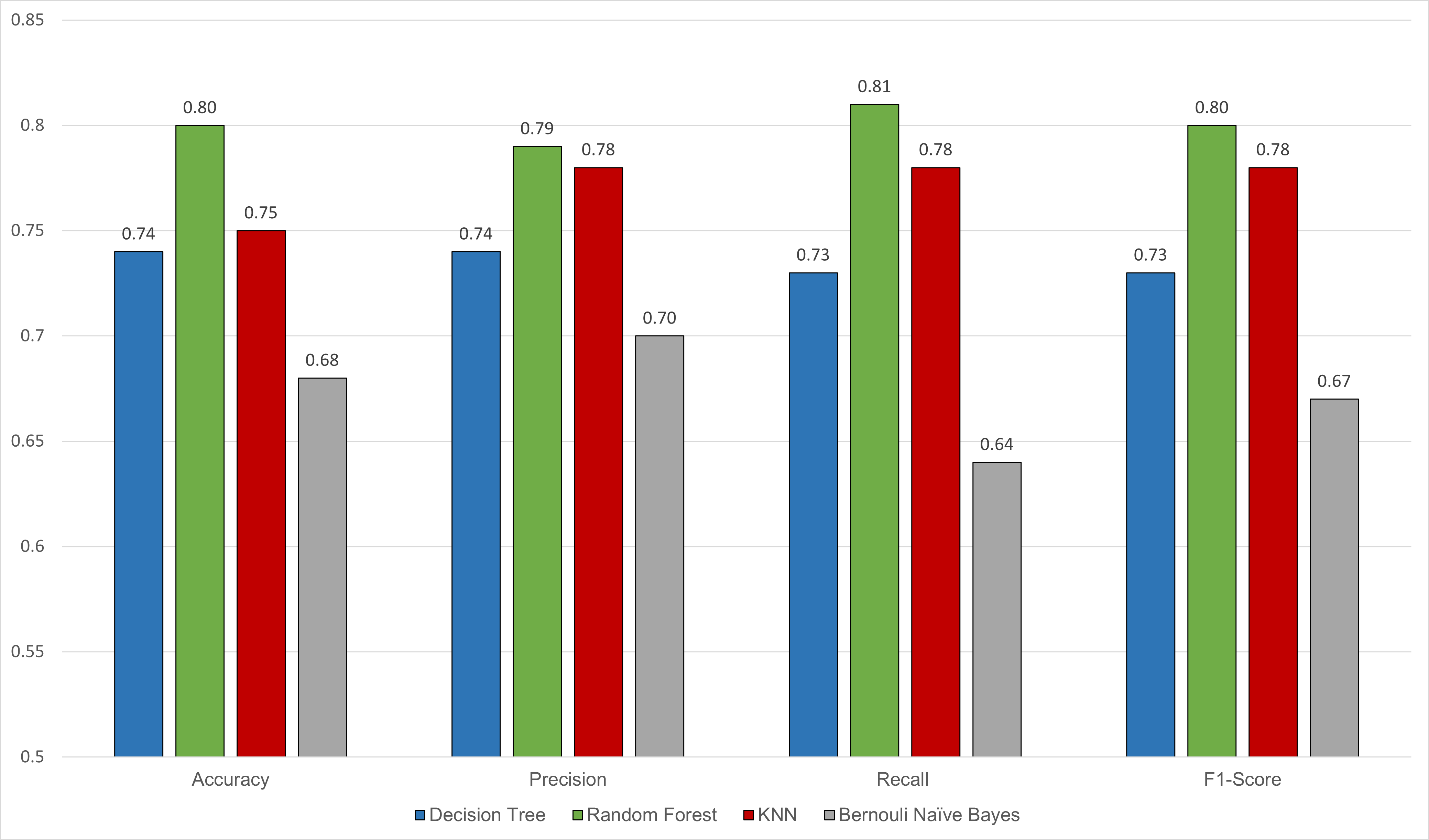}
  \caption{Classification results for Model 2}
  \label{fig:Results2}
\end{figure}

\begin{figure}[htbp]
  \centering
  \includegraphics[width=\linewidth]{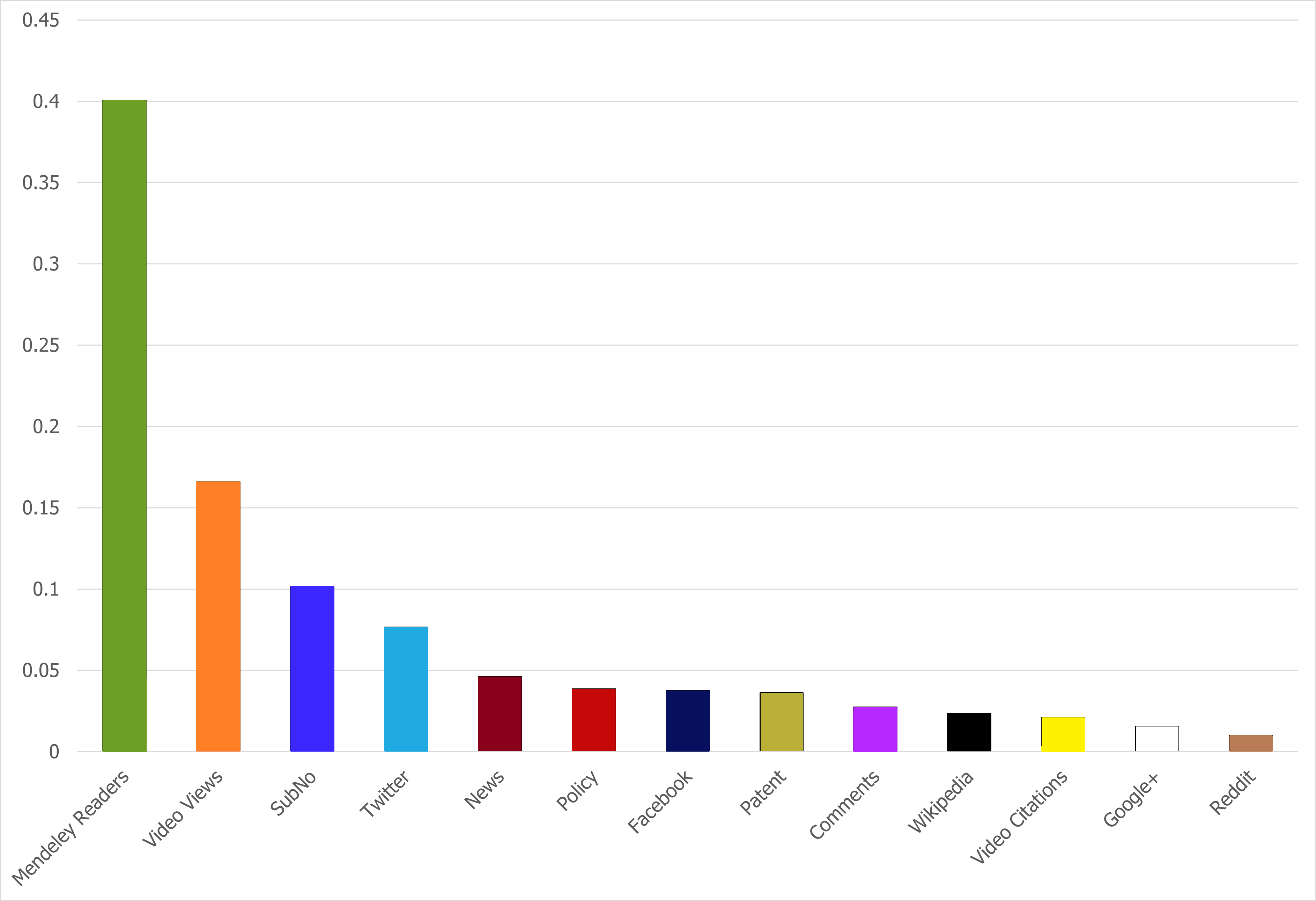}
  \caption{Important features for the Random Forest classifier for Model 2}
  \label{fig:Feature2}
\end{figure}

To predict whether an article will reach a level of popularity or scholarly impact, we built the second model to answer RQ2. For the second model, Random Forest performed the best overall, with an accuracy level and an F1-score of 0.80. KNN performed second best with an accuracy level of 0.75 and an F1-score of 0.78. Figure \ref{fig:Results2} shows the evaluation metrics of the classification models built on dataset C1 to predict the binary target variable. Figure \ref{fig:Feature2} shows feature importance in the model built by the Random Forest classifier. We can see that the average number of Video Views citing a research article is the second-most important feature, after Mendeley Readers, in classifying citations of research articles.

\begin{figure}[htbp]
  \centering
  \includegraphics[width=\linewidth]{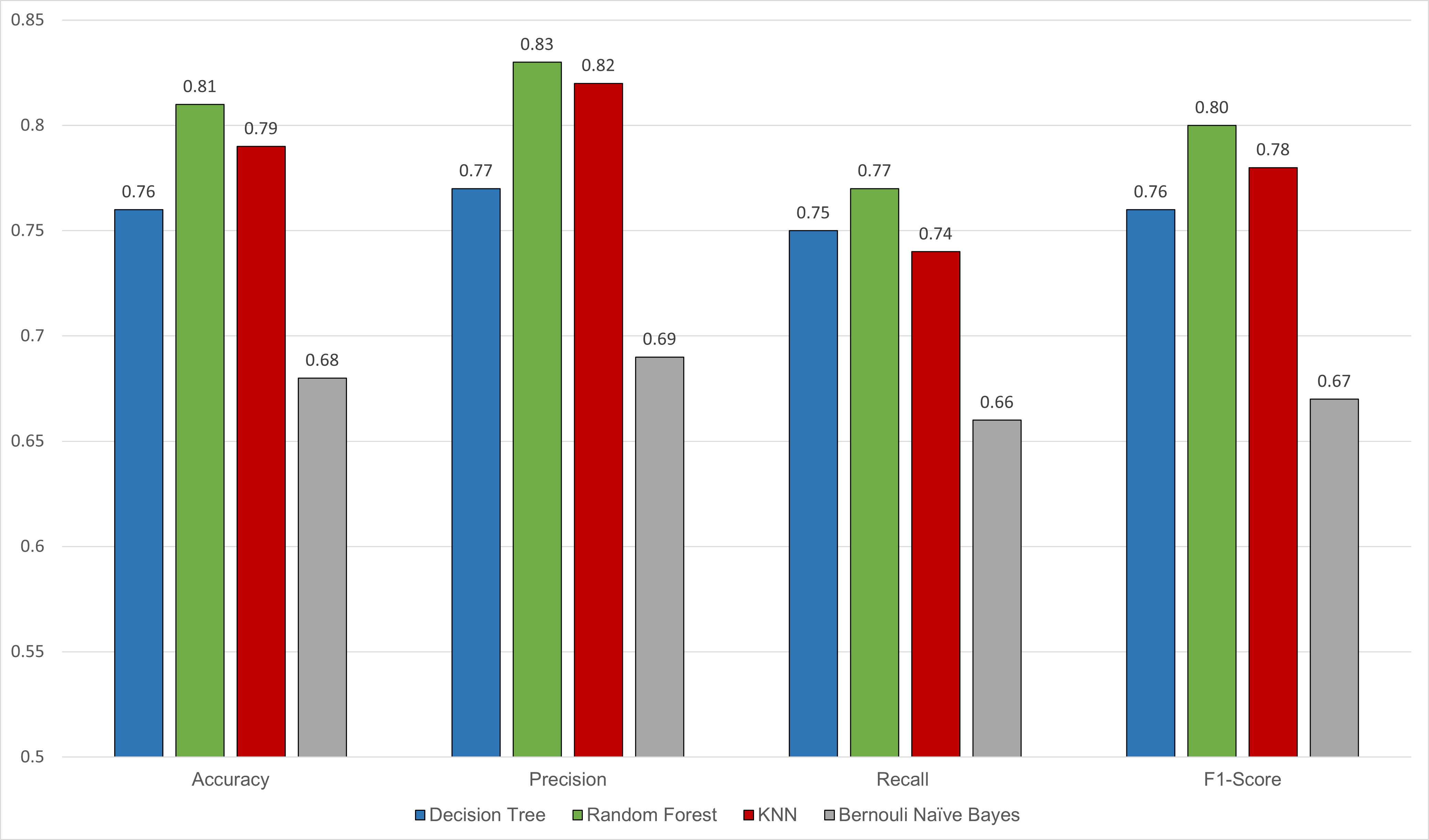}
  \caption{Classification results for Model 3}
  \label{fig:Results3}
\end{figure}

\begin{figure}[htbp]
  \centering
  \includegraphics[width=\linewidth,height=8cm,keepaspectratio]{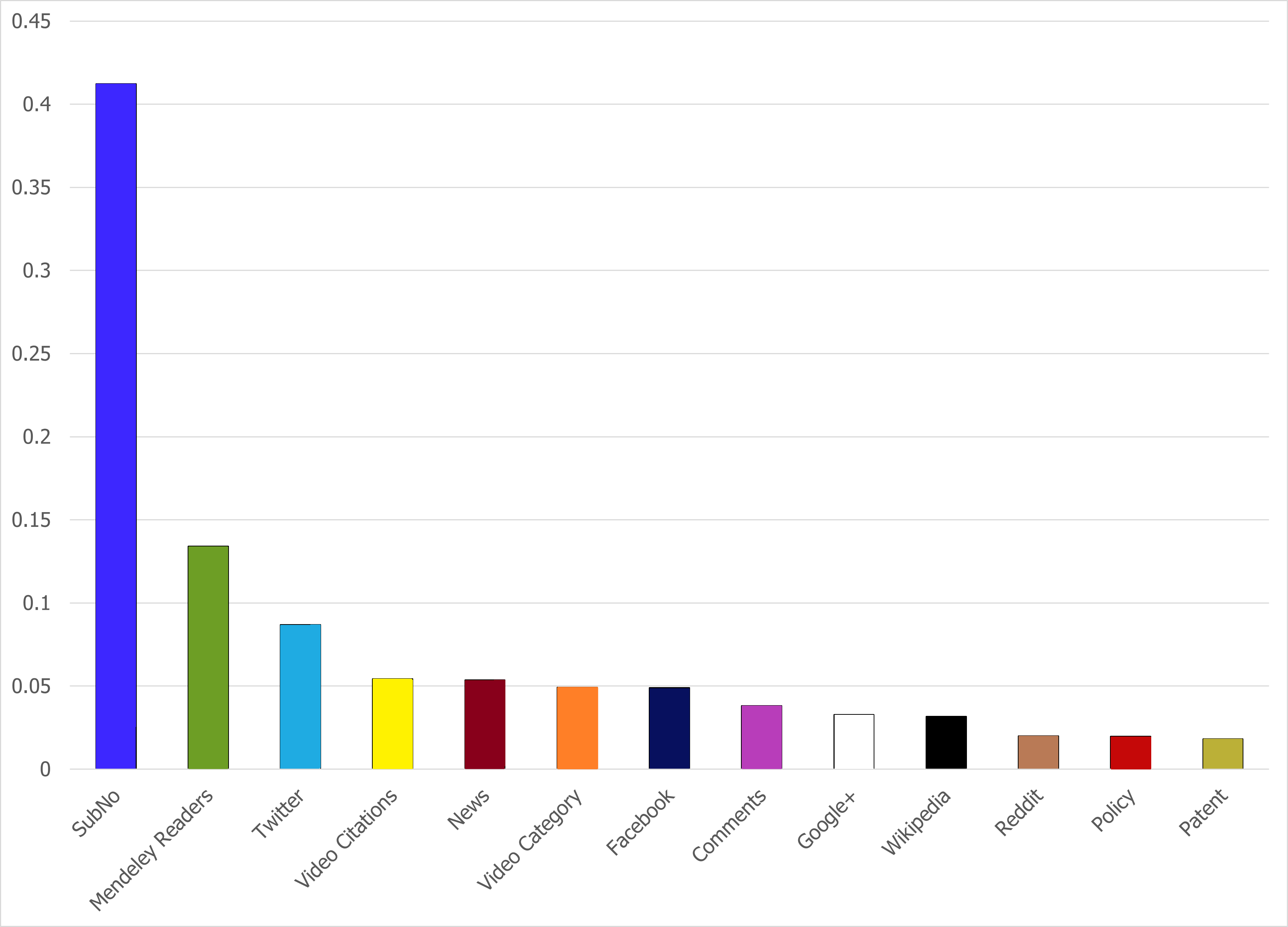}
  \caption{Most important features for the Random Forest classifier for Model 3}
  \label{fig:Feature3}
\end{figure}

We built the third model with the features and the target variable mentioned in Table \ref{tab:overview} to answer RQ2. The third model correlates video popularity and societal impact. The Video Category feature was converted to numerical features using One Hot Encoding. Figure \ref{fig:Results3} shows the results of the classifiers: Random Forest again performed best, achieving an accuracy level of 0.81 and an F1-score of 0.80, and KNN performed second best with an accuracy level of 0.79 and an F1-score of 0.78. Figure \ref{fig:Feature3} shows feature importance in the model built by the Random Forest classifier. SubNo and Mendeley Readers were the most important features in building the model.

\section{Discussion}
Video citations of research articles may help increase the popularity of the articles in the research community and eventually increase citations of research articles. In this study, we analyzed citations of research articles in the descriptions of videos on YouTube. The rise of video citations over recent years, along with the increasing number of conferences requiring a video submission of accepted scholarly research, have paved the way to identify important features that contribute to the popularity and impact of research through the results found in Section 4.1 and 4.2. It is essential to analyze these video citations and the social media mentions of research articles to identify the crucial factors that improve the popularity of videos and research articles. We collected data from YouTube and Altmetric to discover hidden patterns and explore the relationship between research articles and video citations. We combined the datasets, analyzed them using visualization techniques, and built machine learning models to predict important target variables. For the analysis, we examined the combined datasets in relation to research articles and videos, which led us to build three datasets to predict video citations, video views, and citations of research articles.
\par
From our analysis of the large A1 dataset of research articles, we found that around three-quarters of the dataset’s research articles were not cited in videos. In section 4.1, we found that research related to Medicine and Biochemistry received the highest number of video citations. We also found that most of the videos citing research articles were in either the Science \& Technology or the Education category. Videos categorized as Entertainment mentioned fewer research articles but attracted the fourth-highest number of views after Science \& Technology, Education, and People \& Blogs. This result suggests that scientific results are being used for entertainment and educational purposes. Through our analysis of the video dataset B, we found positive correlations between the number of video views, likes, and dislikes, as seen in Figure \ref{fig:datanalysis5}. This is self-evident since users usually view a video before adding a like or dislike. This is also in line with findings reported by \cite{Welbourne2016} and \cite{Pinto2013}, who found that other video metadata such as likes, dislikes, and comment counts are highly correlated with video views. Given these correlations, we removed all these features from consideration, with the exception of the video views feature. However, contrary to previous findings, the comment count did not have a high correlation with views of scientific videos, as seen in Figure \ref{fig:datanalysis5}. This hesitancy to comment publicly on scientific videos could be related to the hesitancy to comment publicly on research articles. Around 80\% of researchers considered that their comments might affect their reputation or how others perceive them \citep{Hemminger2014}.
\par
For our second part of the study, as can be seen in section 4.2, we considered four classification algorithms—Bernoulli Naive Bayes, Random Forest, Decision Tree, and K-Nearest Neighbors—and applied them to each of the three cases to predict three target variables. We evaluated the models based on precision, recall, accuracy, and F1-score. Prediction of whether a research article is cited in a video is the first model of our study for which we used social media mentions of the research article to predict a binary variable indicating video citation. Twitter Mentions and News Mentions of research articles were important factors that contributed to attracting a video citation for a research article. Research articles mentioned in news and tweets cause buzz across the internet and have a high chance of being cited in videos. Citations show the importance and usefulness of a research article and are a strong indication of its popularity. The second model in this study predicted the binary variable, indicating whether a research article had more citations than the median of all the citations of research articles in the dataset. We found that Mendeley Readership and Video Views contributed greatly to this scholarly impact prediction. This result is in line with previous studies in which Mendeley readership is found to be important for predicting citation counts \citep{Thelwall_2018}. However, Video Views are a new feature that plays a vital role in garnering citations, reaching wider audiences, and fostering the popularity and impact of research articles. Our third model predicted Video Views. We found that the subscriber count of a channel, Mendeley Readers, and Twitter Mentions of research articles were important in predicting views of videos citing those research articles. This shows that the number of followers for a YouTube channel plays an essential role in the number of scientific video views. Random Forest achieved the best results for all the models built in this study.

A limitation of this study is that it relied on social media features from a single source Altmetric.com. It also does not consider other related factors, such as the textual content of the research or the video content. Another limitation of our work is that the built models did not consider the temporal nature of the features present in the various datasets, which needs   continuous data collection. We also assumed that citations and views could be considered indicators of scholarly and societal impact, respectively, whereas other factors can also influence these impacts. Moreover, the prediction of the target feature of the models can only be applicable after the accumulation of important social media features, which is a cold-start problem. Furthermore, we used multiple visualizations to explore trends and patterns in our collected datasets. One possible improvement for further studies is to relate and synthesize information from these visualizations \citep{sun2021towards, sun2021sightbi, shaikh}, which may be help gain some more in-depth insights.

Our process offers insight into ways to make improvements in future iterations by incorporating some traditional factors such as the h-index of authors, scholarly venues, and temporal features that have been useful in other studies. Additionally, it would be helpful to distinguish between subjects and research fields to provide a basis for building models for individual areas that would, in turn, improve the results. Furthermore, we considered Dimensions Citations, and it would be worthwhile to check other sources such as Google Scholar. Another direction to explore is the use of research in videos. Do the videos include results drawn principally from the abstract, methods, or results section? Further, the reasons for including references to research articles in videos remain unexplored. For example, do the videos cite research to promote content, provide a message within the video, increase public trust in the content, sell a product, or even spread misinformation? Lastly, our results could be compared with research focused on measuring and predicting other types of societal impact, such as in economic, political, and health areas. 

\section{Conclusion and Future Work}
In this study, we analyzed the impact of YouTube on research by building classification models to predict video citations, scholarly citations, and video views. We found that fewer research articles were cited in videos in a large dataset of research articles. Most of the articles cited in videos belonged to the Medicine and Biochemistry categories. These subjects also have the highest number of video citations in comparison to other subjects. The number of video citations has been increasing in recent years, with most of the videos citing research articles categorized as Science \& Technology, Education, and People \& Blogs. Science \& Technology and Education are the categories of videos that garnered the most views. We found a high-to-mid correlation among views, likes, dislikes, and the subscriber count of videos. We also observed that Random Forest performed the best of all the models built on the datasets. We found that News mentions and Twitter mentions of a research article are important factors in determining a research article’s video citation. Another important finding of the study is that the view count of a video was an important feature in predicting a research article’s citations. For future work, we plan to use more textual and temporal features of the videos and articles. We will study the topics of videos citing research articles and the relationship between the video categories and the research articles’ subjects.

\section*{Acknowledgment}
This research is supported in part by NSF Grants SMA-2022443, IIS-2002082, and the Research and Artistry Opportunity Grant from Northern Illinois University.


%%\begin{acks}
%%This work is supported in part by NSF Grant No. 2022443.
%%\end{acks}

\bibliography{sample-base}

\begin{thebibliography}{}

\bibitem[AAO, 2021]{AAO2021}
AAO (2021).
\newblock Video submission instructions for aao.
\newblock
  \url{https://www.aao.org/annual-meeting/presenter/video-submission-instructions}.
\newblock [Online; accessed 04-April-2021].

\bibitem[Agazio and Buckley, 2009]{Agazio2009}
Agazio, J. and Buckley, K.~M. (2009).
\newblock An untapped resource: using youtube in nursing education.
\newblock {\em Nurse educator}, 34(1):23—28.

\bibitem[Akella et~al., 2021]{AKELLA2021101128}
Akella, A.~P., Alhoori, H., Kondamudi, P.~R., Freeman, C., and Zhou, H. (2021).
\newblock Early indicators of scientific impact: Predicting citations with
  altmetrics.
\newblock {\em Journal of Informetrics}, 15(2):101128.

\bibitem[Alexa, 2021]{AlexaTopSites}
Alexa (2021).
\newblock Alexa - top sites.
\newblock \url{https://www.alexa.com/topsites/}.
\newblock [Online; accessed 01-April-2021].

\bibitem[Beehler and Griffiths, 2020]{Beehler2020}
Beehler, B. and Griffiths, D. (2020).
\newblock Going viral: Taking your conference online for covid-19.
\newblock
  \url{https://www.insidehighered.com/advice/2020/03/16/how-shift-your-conference-online-light-coronavirus-opinion}.
\newblock [Online; accessed 03-April-2021].

\bibitem[Bonifati et~al., 2020]{bonifati2020holding}
Bonifati, A., Guerrini, G., Lutz, C., Martens, W., Mazilu, L., Paton, N.,
  Salles, M. A.~V., Scholl, M.~H., and Zhou, Y. (2020).
\newblock Holding a conference online and live due to covid-19.

\bibitem[Borghol et~al., 2012]{Borghol_2012}
Borghol, Y., Ardon, S., Carlsson, N., Eager, D., and Mahanti, A. (2012).
\newblock The untold story of the clones.
\newblock {\em Proceedings of the 18th ACM SIGKDD international conference on
  Knowledge discovery and data mining - KDD ’12}.

\bibitem[Bornmann, 2014]{BORNMANN2014895}
Bornmann, L. (2014).
\newblock Do altmetrics point to the broader impact of research? an overview of
  benefits and disadvantages of altmetrics.
\newblock {\em Journal of Informetrics}, 8(4):895--903.

\bibitem[Bornmann et~al., 2019]{BORNMANN2019325}
Bornmann, L., Haunschild, R., and Adams, J. (2019).
\newblock Do altmetrics assess societal impact in a comparable way to case
  studies? an empirical test of the convergent validity of altmetrics based on
  data from the uk research excellence framework (ref).
\newblock {\em Journal of Informetrics}, 13(1):325--340.

\bibitem[Brodersen et~al., 2012]{Broderson}
Brodersen, A., Scellato, S., and Wattenhofer, M. (2012).
\newblock Youtube around the world: Geographic popularity of videos.
\newblock In {\em Proceedings of the 21st International Conference on World
  Wide Web}, WWW '12, page 241–250, New York, NY, USA. Association for
  Computing Machinery.

\bibitem[CHI, 2021]{CHI2021}
CHI (2021).
\newblock Technical requirements and guidelines for videos at chi.
\newblock
  \url{https://chi2021.acm.org/for-authors/presenting/papers/technical-requirements-and-guidelines-for-videos-at-chi}.
\newblock [Online; accessed 04-April-2021].

\bibitem[Chtouki et~al., 2012]{Chtouki}
Chtouki, Y., Harroud, H., Khalidi, M., and Bennani, S. (2012).
\newblock The impact of youtube videos on the student's learning.
\newblock In {\em 2012 International Conference on Information Technology Based
  Higher Education and Training (ITHET)}, pages 1--4.

\bibitem[Falk and Hagsten, 2021]{falk2021international}
Falk, M.~T. and Hagsten, E. (2021).
\newblock When international academic conferences go virtual.
\newblock {\em Scientometrics}, 126(1):707--724.

\bibitem[Figueiredo et~al., 2014]{Figueiredo2014}
Figueiredo, F., Almeida, J.~M., Benevenuto, F., and Gummadi, K.~P. (2014).
\newblock Does content determine information popularity in social media? a case
  study of youtube videos' content and their popularity.
\newblock In {\em Proceedings of the SIGCHI Conference on Human Factors in
  Computing Systems}, CHI '14, page 979–982, New York, NY, USA. Association
  for Computing Machinery.

\bibitem[Freeman et~al., 2020]{cole-2020-group}
Freeman, C., Alhoori, H., and Shahzad, M. (2020).
\newblock Measuring the diversity of facebook reactions to research.
\newblock {\em Proc. ACM Hum.-Comput. Interact.}, 4(GROUP).

\bibitem[Freeman et~al., 2019]{cole-2019-jcdl}
Freeman, C., Roy, M.~K., Fattoruso, M., and Alhoori, H. (2019).
\newblock Shared feelings: Understanding {Facebook} reactions to scholarly
  articles.
\newblock In {\em 2019 ACM/IEEE Joint Conference on Digital Libraries (JCDL)},
  pages 301--304.

\bibitem[Hemminger and TerMaat, 2014]{Hemminger2014}
Hemminger, B.~M. and TerMaat, J. (2014).
\newblock Annotating for the world: Attitudes toward sharing scholarly
  annotations.
\newblock {\em Journal of the Association for Information Science and
  Technology}, 65(11):2278--2292.

\bibitem[Hovden, 2013]{Hovden}
Hovden, R. (2013).
\newblock Bibliometrics for internet media: Applying the h-index to youtube.
\newblock {\em Journal of the American Society for Information Science and
  Technology}, 64(11):2326--2331.

\bibitem[Jaffar, 2012]{Jaffar2012}
Jaffar, A.~A. (2012).
\newblock Youtube: An emerging tool in anatomy education.
\newblock {\em Anatomical Sciences Education}, 5(3):158--164.

\bibitem[Johnston et~al., 2018]{JOHNSTON2018151}
Johnston, A.~N., Barton, M.~J., Williams-Pritchard, G.~A., and Todorovic, M.
  (2018).
\newblock Youtube for millennial nursing students; using internet technology to
  support student engagement with bioscience.
\newblock {\em Nurse Education in Practice}, 31:151--155.

\bibitem[June et~al., 2014]{June2014}
June, S., Yaacob, A., and Kheng, Y.~K. (2014).
\newblock Assessing the use of youtube videos and interactive activities as a
  critical thinking stimulator for tertiary students: An action research.
\newblock {\em International Education Studies}, 7:56--67.

\bibitem[Khan and Vong, 2014]{Khan2014}
Khan, G. and Vong, S. (2014).
\newblock Virality over youtube: an empirical analysis.
\newblock {\em Internet Res.}, 24:629--647.

\bibitem[Kousha et~al., 2022]{Thelwall-covid}
Kousha, K., Thelwall, M., and Bickley, M. (2022).
\newblock The high scholarly value of grey literature before and during
  covid-19.
\newblock {\em Scientometrics}, 127(6):3489--3504.

\bibitem[Liu et~al., 2022]{Bu}
Liu, M., Bu, Y., Chen, C., Xu, J., Li, D., Leng, Y., Freeman, R.~B., Meyer,
  E.~T., Yoon, W., Sung, M., Jeong, M., Lee, J., Kang, J., Min, C., Song, M.,
  Zhai, Y., and Ding, Y. (2022).
\newblock Pandemics are catalysts of scientific novelty: Evidence from
  covid-19.
\newblock {\em Journal of the Association for Information Science and
  Technology}, 73(8):1065--1078.

\bibitem[Ma et~al., 2017]{Ma2017}
Ma, C., Yan, Z., and Chen, C.~W. (2017).
\newblock Larm: A lifetime aware regression model for predicting youtube video
  popularity.
\newblock In {\em Proceedings of the 2017 ACM on Conference on Information and
  Knowledge Management}, CIKM '17, page 467–476, New York, NY, USA.
  Association for Computing Machinery.

\bibitem[Madathil et~al., 2015]{Madathil}
Madathil, K.~C., Rivera-Rodriguez, A.~J., Greenstein, J.~S., and Gramopadhye,
  A.~K. (2015).
\newblock Healthcare information on youtube: A systematic review.
\newblock {\em Health Informatics Journal}, 21(3):173--194.
\newblock PMID: 24670899.

\bibitem[Pedregosa et~al., 2011]{pedregosa11a}
Pedregosa, F., Varoquaux, G., Gramfort, A., Michel, V., Thirion, B., Grisel,
  O., Blondel, M., Prettenhofer, P., Weiss, R., Dubourg, V., Vanderplas, J.,
  Passos, A., Cournapeau, D., Brucher, M., Perrot, M., and {{\'E}}douard
  Duchesnay (2011).
\newblock Scikit-learn: Machine learning in python.
\newblock {\em Journal of Machine Learning Research}, 12(85):2825--2830.

\bibitem[Pinto et~al., 2013]{Pinto2013}
Pinto, H., Almeida, J.~M., and Gon\c{c}alves, M.~A. (2013).
\newblock Using early view patterns to predict the popularity of youtube
  videos.
\newblock In {\em Proceedings of the Sixth ACM International Conference on Web
  Search and Data Mining}, WSDM '13, page 365–374, New York, NY, USA.
  Association for Computing Machinery.

\bibitem[Price, 2020]{MichaelPrice}
Price, M. (2020).
\newblock As covid-19 forces conferences online, scientists discover upsides of
  virtual format.
\newblock
  \url{https://www.sciencemag.org/careers/2020/04/covid-19-forces-conferences-online-scientists-discover-upsides-virtual-format}.
\newblock [Online; accessed 03-April-2021].

\bibitem[Rodriguez, 2021]{YouTubepop2}
Rodriguez, S. (2021).
\newblock Youtube is social media’s big winner during the pandemic.
\newblock
  \url{https://www.cnbc.com/2021/04/07/youtube-is-social-medias-big-winner-during-the-pandemic.html}.
\newblock [Online; accessed 06-June-2021].

\bibitem[Shahzad and Alhoori, 2022]{Murtuza2022a}
Shahzad, M. and Alhoori, H. (2022).
\newblock Public reaction to scientific research via twitter sentiment
  prediction.
\newblock {\em Journal of Data and Information Science}, 7(1):97--124.

\bibitem[Shahzad et~al., 2022]{Murtuza2022b}
Shahzad, M., Alhoori, H., Freedman, R., and Rahman, S.~A. (2022).
\newblock Quantifying the online long-term interest in research.
\newblock {\em Journal of Informetrics}, 16(2):101288.

\bibitem[Shaikh and Alhoori, 2019]{Abdul-jcdl-2019}
Shaikh, A.~R. and Alhoori, H. (2019).
\newblock Predicting patent citations to measure economic impact of scholarly
  research.
\newblock In {\em 2019 ACM/IEEE Joint Conference on Digital Libraries (JCDL)},
  pages 400--401.

\bibitem[Shaikh et~al., 2022]{shaikh}
Shaikh, A.~R., Koop, D., Alhoori, H., and Sun, M. (2022).
\newblock Toward systematic design considerations of organizing multiple views.
\newblock {\em 2022 IEEE VIS conference}.

\bibitem[Snelson et~al., 2012]{Snelson2012}
Snelson, C., Rice, K., and Wyzard, C. (2012).
\newblock Research priorities for youtube and video-sharing technologies: A
  delphi study.
\newblock {\em British Journal of Educational Technology}, 43(1):119--129.

\bibitem[Suciu, 2021]{YouTubepop}
Suciu, P. (2021).
\newblock Youtube remains the most dominant social media platform.
\newblock
  \url{https://www.forbes.com/sites/petersuciu/2021/04/07/youtube-remains-the-most-dominant-social-media-platform/?sh=284719d36322}.
\newblock [Online; accessed 05-June-2021].

\bibitem[Sud and Thelwall, 2014]{sud2014evaluating}
Sud, P. and Thelwall, M. (2014).
\newblock Evaluating altmetrics.
\newblock {\em Scientometrics}, 98(2):1131--1143.

\bibitem[Sugimoto et~al., 2017]{Cassidy}
Sugimoto, C.~R., Work, S., LariviÃ¨re, V., and Haustein, S. (2017).
\newblock Scholarly use of social media and altmetrics: A review of the
  literature.
\newblock {\em Journal of the Association for Information Science and
  Technology}, 68(9):2037--2062.

\bibitem[Sun et~al., 2021a]{sun2021towards}
Sun, M., Namburi, A., Koop, D., Zhao, J., Li, T., and Chung, H. (2021a).
\newblock Towards systematic design considerations for visualizing cross-view
  data relationships.
\newblock {\em IEEE Transactions on Visualization and Computer Graphics}.

\bibitem[Sun et~al., 2021b]{sun2021sightbi}
Sun, M., Shaikh, A.~R., Alhoori, H., and Zhao, J. (2021b).
\newblock Sightbi: Exploring cross-view data relationships with biclusters.
\newblock {\em IEEE Transactions on Visualization and Computer Graphics},
  28(1):54--64.

\bibitem[Susarla et~al., 2012]{Susarla2012}
Susarla, A., Oh, J.-H., and Tan, Y. (2012).
\newblock Social networks and the diffusion of user-generated content: Evidence
  from youtube.
\newblock {\em Information Systems Research}, 23(1):23--41.

\bibitem[TechPostPlus, 2019]{TechPost}
TechPostPlus (2019).
\newblock Youtube video categories list.
\newblock
  \url{https://techpostplus.com/youtube-video-categories-list-faqs-and-solutions/}.
\newblock [Online; accessed 05-April-2021].

\bibitem[Thelwall and Nevill, 2018]{Thelwall_2018}
Thelwall, M. and Nevill, T. (2018).
\newblock Could scientists use altmetric.com scores to predict longer term
  citation counts?
\newblock {\em Journal of Informetrics}, 12(1):237–248.

\bibitem[{Trzciński} and {Rokita}, 2017]{Trzcinski2017}
{Trzciński}, T. and {Rokita}, P. (2017).
\newblock Predicting popularity of online videos using support vector
  regression.
\newblock {\em IEEE Transactions on Multimedia}, 19(11):2561--2570.

\bibitem[Welbourne and Grant, 2016]{Welbourne2016}
Welbourne, D.~J. and Grant, W.~J. (2016).
\newblock Science communication on youtube: Factors that affect channel and
  video popularity.
\newblock {\em Public Understanding of Science}, 25(6):706--718.
\newblock PMID: 25698225.

\bibitem[Weller et~al., 2015]{Weller2015}
Weller, Katrin.~Welpe, I.~M., Wollersheim, J., Ringelhan, S., and Osterloh, M.
  (2015).
\newblock {\em Social Media and Altmetrics: An Overview of Current Alternative
  Approaches to Measuring Scholarly Impact}, pages 261--276.
\newblock Springer International Publishing, Cham.

\bibitem[YouTube, 2021]{YouTubePress}
YouTube (2021).
\newblock Press - youtube.
\newblock \url{https://www.youtube.com/about/press/}.
\newblock [Online; accessed 01-April-2021].

\bibitem[Yu et~al., 2015]{Yu2015}
Yu, H., Xie, L., and Sanner, S. (2015).
\newblock The lifecyle of a youtube video: Phases, content and popularity.
\newblock In {\em ICWSM}.

\end{thebibliography}

%% End of file `sample-acmsmall.tex'.

\end{document}